*Retrospective of the ARPA-E ALPHA fusion program*

C. L. Nehl,[1] R. J. Umstattd,[2,*] W. R. Regan,[2,**] S. C. Hsu,[2,†] P. B. McGrath[2]

1: Booz Allen Hamilton, McLean, VA 22102
2: Advanced Research Projects Agency–Energy, U.S. Department of Energy, Washington, DC 20585
* R. J. Umstattd is presently affiliated with Fusion Industry Association, Gaithersburg, MD 20879
** W. R. Regan is presently affiliated with X, the Moonshot Factory, Mountain View, CA 94043
†Corresponding author email:  scott.hsu@hq.doe.gov

*Abstract*

This paper provides a retrospective of the ALPHA (Accelerating Low-cost Plasma Heating and Assembly) fusion program of the Advanced Research Projects Agency–Energy (ARPA-E) of the U.S. Department of Energy.  ALPHA's objective was to catalyze research and development efforts to enable substantially lower-cost pathways to economical fusion power.  To do this in a targeted, focused program, ALPHA focused on advancing the science and technology of pulsed, intermediate-density fusion approaches, including magneto-inertial fusion and Z-pinch variants, that have the potential to scale to commercially viable fusion power plants.  The paper includes a discussion of the origins and framing of the ALPHA program, a summary of project status and outcomes, a description of associated technology-transition activities, and thoughts on a potential follow-on ARPA-E fusion program.

*Introduction:*

In 2014 the Advanced Research Projects Agency—Energy (ARPA-E) of the U.S. Department of Energy (DOE) launched a new research program on low-cost approaches to fusion-energy development[1].  The "Accelerating Low-Cost Plasma Heating and Assembly" (ALPHA) program set out to enable more rapid progress towards fusion energy by establishing a wider range of technological options that could be pursued with smaller, lower-cost experiments, short development and construction times, and high experimental throughput.   Mainstream fusion research generally refers to magnetically or inertially confined fusion, both of which require expensive facilities for reasons briefly described below and explored in more detail in several books[2][3]. ALPHA focused on magneto-inertial fusion (MIF), a class of pulsed fusion approaches with fuel densities in between those of magnetic and inertial fusion[4], [5] [6].  This paper presents a brief background on the origins of the ALPHA program, the results achieved by ALPHA-funded teams, and a look ahead to potential next steps for low-cost fusion development.

*Origins*

ARPA-E's mission is to develop transformational new energy technologies[7].  While DOE has pursued fusion energy as a potentially transformational opportunity for decades, ARPA-E had not supported any work in fusion prior to the ALPHA program.  This was in part due to a perception that fusion was inherently the realm of "big science" and that ARPA-E, which runs relatively small, targeted, short-term programs across a wide spectrum of energy technologies—did not have a role to play in that development.  In launching ALPHA, ARPA-E sought to change this dynamic and bring new players into the field – both in terms of the kinds of teams doing fusion development (e.g., smaller groups and private startups), and in terms of the sources of funding (e.g., private investors). The ALPHA program was also a way for ARPA-E to address a longer-term problem in energy development with a targeted program. The motivation and timing for the ALPHA program were driven by three major factors:

1) analysis suggesting the potential for lower-cost pathways with fuel densitines between those of the mainstream approaches of magnetic confinement fusion (MCF) and inertial confinement fusion (ICF) [8], [9],
2) significant experimental results from magnetized inertial confinement fusion [10] and from the Magnetized Liner Inertial Fusion (MagLIF)[11] program that supported this analysis, and
3) growing private sector investment in fusion[12][13], opening an opportunity for new approaches if they can achieve performance gains at costs compatible with private investors.

*1: Potential for lower-cost pathways:* The overwhelming majority of fusion research funding is currently devoted to major programs in MCF (principally the ITER collaboration and supporting plasma science in conventional tokamaks), and in ICF (principally laser-driven systems such as the National Ignition Facility, NIF, in the U.S.). ITER and NIF are each multi-billion dollar facilities, and the costs are driven in large part by performance requirements that are intrinsic to their respective approaches. ITER, which will operate as a long-pulse device with an ion density of approximately $10^{14}$ cm$^{-3}$, requires an exceptionally large vacuum vessel and magnet set to contain a plasma of sufficient size to meet and exceed Lawson conditions, and the costs of the vacuum vessel and magnets are correspondingly large[14]. NIF, a pulsed device that compresses targets to ion densities greater than $10^{26}$ cm$^{-3}$, requires exceptionally high power and power density to overcome the thermal losses (hundreds of TW peak power), and the cost of a MJ-class laser and optics systems to deliver sufficient energy in a sufficiently short time to the target system drives high costs for the machine[15][16]. ITER and NIF are the leading facilities within MCF and ICF, respectively. For the purposes of burning plasma research (as in ITER) or ignition and propagating burn (as in NIF), these "big science" projects have arguably the lowest scientific risk for achieving their respective goals. However, there are a wide range of alternative approaches spanning the full range of parameter space for fusion plasmas, including many that lie near the middle of the ten-plus orders of magnitude in ion density between MCF and ICF[8]. In fact, there are a number of analyses suggesting that some of these intermediate-density approaches with very high magnetic fields (megagauss or higher) may be able to achieve Lawson conditions at significantly lower costs than the mainline MCF or ICF approaches. The reasoning behind these analyses varies – from an optimal balance between the minimum size/energy of a plasma against the minimum power to overcome thermal losses[8], to an optimum magnetic field in the megagauss range for sizing plasma and pulsed power components[9], to the power density of the fusion core matched to practical reactor scaling [17], but they each suggest that the space in between ITER and NIF may be less costly to explore than the MCF and ICF extremes.

*2. MagLIF experiments constituting proof-of-concept for MIF:* The analyses referenced above have developed over decades, but there has been relatively little exploration of the concepts that might fall in this range, and thus little experimental data or validated models to offer more detailed support. However, in 2014, Sandia completed their first integrated shots of the MagLIF experiments, which used the Z-machine to implode a pre-heated and magnetized cylindrical D plasma target [18]. The experiments reached peak ion densities exceeding $10^{22}$ cm$^{-3}$ and multi-keV temperatures, producing significant DD neutron yields from thermonuclear fusion[11]. These results – which came very early in the first campaigns, and have subsequently been exceeded – represent the first significant experimental evidence to support the claim that intermediate-density, magnetized fusion approaches could be significantly lower in cost that MCF or ICF [19]. The MagLIF experiments were performed on the Z-machine, a pulsed power machine that was not built nor optimized with MagLIF in mind. The Z-machine is more than an order of magnitude lower in cost than NIF, which was designed from the beginning with laser indirect-drive fusion as its primary mission [20].  While the MagLIF results do not represent a new record in fusion yield, the very fact that these experiments produced high yield in early experiments on a non-purpose built, relatively low-cost machine suggests that this is an area of fusion research that warrants further exploration.

*3. Increased private interest in fusion:* At the same time that these scientific developments were taking place, there was also a growing movement of private investors taking increased interest, and devoting significant private resources, to fusion development.  In the years leading up to the ALPHA program, hundreds of millions of dollars were invested into private fusion companies, led by Tri Alpha Energy (now TAE Technologies) in the U.S., Tokamak Energy in the U.K., and General Fusion in Canada[12], [13]. Acknowledging the extremely high technical risks and long timelines associated with fusion, the interest and appetite for private investors to participate in fusion development signals an opportunity to bring in new players and expand the field.  A central thesis of the ALPHA program was that we could expand the field if we could offer more options for fusion development that could be developed at funding levels compatible with private investment (i.e., tens to hundreds of millions of dollars for R&D, not several billion for scientific proof of concept).  When combined with (1) compelling arguments for low-cost pathways and (2) experimental evidence supporting those arguments, ARPA-E determined that this could offer transformational opportunity to change the trajectory for the field.

Based on this combination of factors, ARPA-E launched the ALPHA program to explore "intermediate-density" fusion approaches with peak ion densities ranging from $10^{18}$-$10^{23}$ cm$^{-3}$[1]. This is a range that includes a diversity of approaches, but all share the common attributes of a magnetized plasma (a "target") that must be compressed in a pulsed fashion (using a "driver") to reach high density and temperature.  Some examples include magneto-inertial fusion (MIF, sometimes called magnetized target fusion, MTF), which utilize an imploding conductive liner to compress a fusion plasma, and stabilized dense Z-pinches, which use direct pulsed power to assemble, compress, and heat a column of fusion fuel.  The focus on intermediate-density approaches reflected the opportunity for low-development cost in a range that was relatively under-explored as compared to MCF and ICF approaches.

Beyond the focus on approaches in the intermediate ion density range, the ALPHA program set specific goals for the cost (<$0.05/MJ delivered driver energy, measured over full driver life), engineering gain (>5 for product of driver efficiency and projected fusion gain), and shot rate (hundreds of shots in ALPHA program, path to >1 Hz operation) of the proposed plasma systems, all with the purpose of achieving rapid experimental progress in the near term, and enabling economical fusion power reactors in the long term[1].  These constraints ruled out many destructive experimental approaches that, such as the use of explosives for compression.  Recent progress in pulsed power technology, such as the continued development of wide bandgap devices for high current/high voltage solid-state switches[21][22], [23] and linear transformer drivers (LTD)[24], [25] offer promise that pulsed fusion approaches can achieve high efficiency, low cost, and high repetition rate.  ALPHA teams were permitted to use "legacy" pulsed power machines to demonstrate performance, but each had to justify that the current and voltage levels, and the required timescales and profiles for discharge could be compatible with eventual efficient operation at high repetition rate (e.g., 1 Hz).

Out of this competitive solicitation, a portfolio consisting of nine teams was selected for award in the ALPHA program.  There was a diversity of approaches within the program, including pulsed magnetic compression, MIF with piston-driven liquid liner compression, MIF with high velocity plasma jet compression, and stabilized Z-pinches.  There were also projects in the portfolio performing applied scientific studies relevant to the densities, magnetic fields, and plasma/liner interface environments encountered in this range of fusion parameter space.  The program also included some exploratory efforts on new components that could be broadly enabling for new fusion concepts.  For the purposes of discussion, we group those projects as "Integrated Concept" teams, which developed integrated plasma and compression systems to produce thermonuclear fusion plasmas; "Driver" teams that developed technologies for liner compression systems that could be applied to MIF concepts (but did not integrate the drivers with plasma targets during the ALPHA program); "Applied Science" teams that performed experimental and simulation studies to better inform intermediate density fusion plasmas and MIF; and "Exploratory Concepts" that developed novel plasma configurations and driver components.  (Note that these groupings were not categories of the FOA, but are rather post-hoc descriptions of the general thrusts within the program.  There is some overlap within these groupings, insofar as all teams performed some level of exploratory work and applied science.) The following section describes the goals, progress, and status for each of the individual projects.

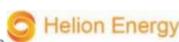

Figure 1:  List of lead organizations and approaches selected in the ALPHA program.

***ALPHA projects and results***

*i. Integrated Concept Teams*

*University of Washington/Lawrence Livermore National Lab: Sheared-flow Z-Pinch for Fusion*

The University of Washington (UW), along with its partner Lawrence Livermore National Laboratory (LLNL), developed a variant of the Z-pinch that exploits sheared flow in the axial direction to mitigate the m=0 and m=1 instabilities that plague Z-pinch plasmas. The concept builds upon prior work in the ZAP and ZAP-HD experiments which demonstrated that a Z-pinch initiated from a high velocity plasma gun experiences a shear flow from r=0 at the center of the Z-pinch axis to r=R at the plasma edge[26], [27][28]. In those experiments, it was shown that at sufficiently high velocities (typically observed as ~10% of the Alfvén velocity times *k*, the axial wave number) from the plasma gun, the shear in the plasma was able to suppress the growth of sausage and kink modes in a Z-pinch over a stable period about 700X the expected instability growth time in a non-sheared Z-pinch, at densities of $10^{16}$-$10^{17}$ cm$^{-3}$ and temperatures of 50-80 eV[27]. Under the ALPHA program, the UW/LLNL team sought to determine if this shear stabilization mechanism scales to fusion conditions, specifically by pushing the current to 100's of kA, thereby increasing the density to ~$10^{17}$ cm$^{-3}$ and temperature to ~1-2 keV. As shown in figure 2, and as summarized in a recent paper, the team was able to demonstrate experimentally that the sheared-flow Z-pinch at high currents (approximately 200 kA) exhibited stability for 5-20 μs, several orders of magnitude longer than the characteristic growth time for sausage and kink instabilities[29]. At these currents, plasmas of 20% Deuterium/80% Hydrogen reached densities of $10^{17}$ cm$^{-3}$ and temperatures estimated at 500 eV-1 keV, and reproducibly generated neutron yields >$10^5$ for 5-μs periods, and observed a scaling of neutron emission with the square of the deuterium ion number density, which suggests thermonuclear origin[29].

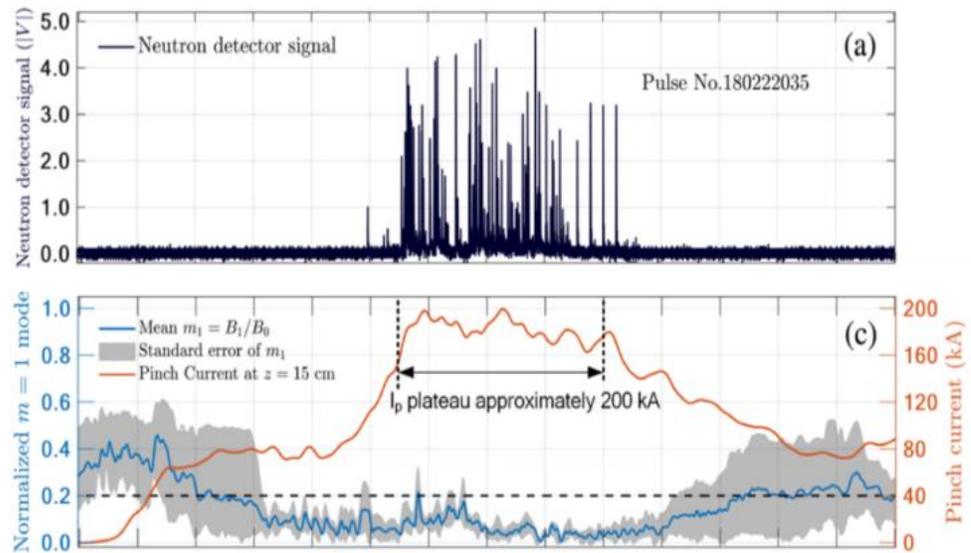

Figure 2: Results from the University of Washington Sheared flow Z-pinch. Top: Signal observed on the scintillator detector, which shows neutron producing during the stable (quiescent) period. Bottom: Normalized magnetic field fluctuation amplitude for the m=1 mode, as measured a multiple locations, which shows that the plasma is relative stable for about 5 uS. Note that the quiescent period aligns to the time in whch neutrons were detected. Figure adapted from PRL 122, 135001

The UW/LLNL team also performed extensive MHD and PIC simulations of the sheared-flow Z-pinch system. As the system pushes to higher currents, temperatures, and densities, the plasma will approach kinetic conditions, and at the outset of the project it remained an open question as to whether the shear stabilization demonstrated in the ZAP and ZAP-HD experiments would hold in the kinetic regime. PIC simulations from LLNL suggest that the shear stabilization mechanism will remain effective at the high currents projected for fusion conditions. [30] [31] This is a valuable addition to the existing literature on sheared flow Z-pinch stabilization. To review it was theoretically predicted that the kink (*m*=1) mode would be stabilized when when the flow shear ($\partial V_z/\partial r$) exceeds $0.1kV_A$ (0.1 times the axial wave number times the Alfvén wave velocity) [26]. This prediction was later verified experimentally, which set the stage for further development of the sheared-flow Z-pinch[32]. Fully kinetic PIC simulations have also shown the suppression of instabilities in sheared flow stabilized Z-pinch plasmas at scales ranging from current experiments up to reactor-scale[30].

Based upon the promising results of the research under the ALPHA program, the team from UW launched a new company, Zap Energy, and has won a follow-on award from ARPA-E to push the sheared-flow Z-pinch to higher currents, possibly necessitating improved materials and designs for high-current-density electrodes, and refinement of timing and current profiles for plasma initiation and for stabilization at increased densities and temperatures[33].

*Helion Energy – Magnetic Compression of Field Reversed Configuration (FRC) Targets for Fusion*

Helion Energy is developing an MIF concept for the compression of an FRC plasma using a high field compression coil.. The concept builds upon prior work at UW and at MSNW LLC, and utilizes the dynamic formation of two FRCs, accelerated towards each other and merged in a central chamber[34]–[36]. A high power magnet coil surrounding the central chamber compresses and heats the FRC. In prior experiments, such as the "Grande" experiment at MSNW, the merged and compressed FRCs reached high temperatures and densities. Based upon empirical scaling relationships from the LSX experiments at UW in the 1990's[36], [37], the Helion team has projected that fusion conditions are achievable in a relatively low-cost machine with increased trapped flux in the FRCs, and increased peak B-field from compression [38]. However, these projections from the relatively low density, steady state conditions of LSX must extrapolate to several orders of magnitude higher in ion density, as well as significant increases in other experimental parameters, and there is limited theoretical or simulation-based understanding of the FRC to confidently make those projections. The ALPHA-supported research sought to increase the trapped flux in the FRCs by a factor of 2 and then compress it to a higher peak magnetic field in order to provide experimental data in the higher density regime, as well as an experimental basis for performance projections for compressed FRC targets in fusion conditions. The experiments (in keeping with prior nomenclature, the updated machine was named "Venti"), proved to be challenging, particularly the mechanical structure of the high-field compression coil, and for keeping the highly compressed FRCs on-axis in the relatively small radius of the central chamber. Even with the aggressive experimental goals, the team was able to conduct over 900 FRC compression shots, and the team observed DD fusion neutrons. As the team reported to an independent review team (JASON) in 2018, Helion's integrated system achieved a density of $8 \times 10^{16}$ ions/cm$^3$, a final magnetic field of 8 T, a final radius of 6 cm, and an energy confinement time at maximum compression of 4 x 10$^{-5}$ s [39]. While this data is encouraging, further experimental measurements of the plasma parameters are necessary to firmly validate these claims. Significantly, the team believes they showed that the micro-scale confinement and macro-scale stability scale as expected. Separate from the ALPHA award, Helion is also pursuing increased performance for a larger scale compressed FRC system. Beyond their technical progress, Helion has also successfully secured private investment (some of it prior to their ALPHA award).

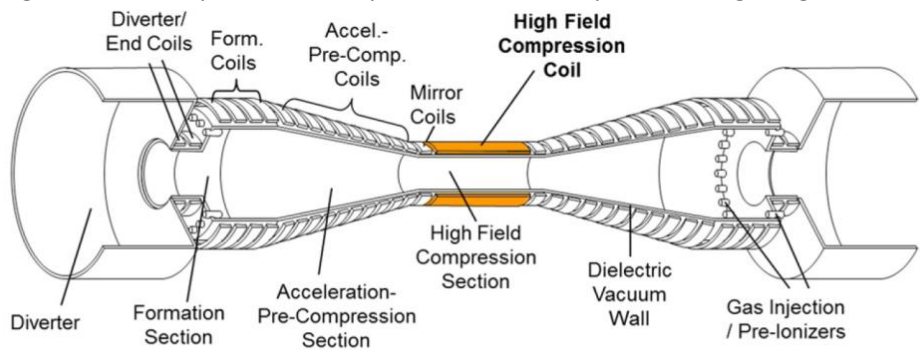

*Figure 3: An early depiction of the Helion approach. Image reproduced from Nucl. Fusion 51 (2011) 053008*

*Magneto-Inertial Fusion Technologies, Inc. (MIFTI)/University of California, San Diego (UCSD)/University of Nevado, Reno (UNR): Staged Z-Pinch Target For Fusion*

MIFTI, along with partner UCSD, is developing the "Staged Z-Pinch" MIF concept, which delivers high current pulsed power to an annular shell of high atomic number gas (e.g., Ar or Kr), which then compresses at high velocity on a cylindrical target of magnetized D-D fuel[40][41] [42] [43]. The geometry of the Staged Z-Pinch is very similar to Sandia's MagLIF, in that both are cylindrical magnetized plasma targets, and are compressed radially by a high-Z liner. However, there are important differences for MIFTI's Staged Z-Pinch concept. First, there is no laser pre-heat of the MIFTI target, instead the pre-compression heating in the plasma target is provided by the initial shock of the imploding high-Z liner at the interface with the (stagnant) low-Z fuel target. Another key difference is that the liner in the Staged Z-Pinch is an annular gas puff, as opposed to a solid metal liner as in MagLIF. The gas puff allows for more rapid experimentation, as each shot does not require replacement of the liner hardware, and in the course of the ALPHA project, MIFTI routinely complete 10's of shots per day (albeit at current levels >10x lower than MagLIF shots on Sandia's Z-machine).  The Staged Z-Pinch concept has been described in a number of simulation-based studies, and was explored in limited experimental work at UC Irvine in the 1990s.  The goals for the ALPHA project were to demonstrate the Staged Z-Pinch at fusion conditions utilizing the 2 TW  (up to 1.2 MA) Zebra pulsed power machine at the Nevada Terawatt Facility at The University of Nevada Reno[44].  Hundreds of shots were completed, exploring a range of parameters for initial magnetization in the target, ion density (both in the target and in the liner), and using different liner species (principally Ar and Kr).  Through the course of several campaigns on Zebra, the MIFTI team was able to produce  consistent and repeatable shots with neutron yields exceeding $10^9$, with top-performing shots (Kr liner imploding on target with initial axial B-field of 10 kG) exceeding $10^{10}$ neutrons[45][42] . The neutron yields appeared to have an isotropic and repeatable distribution suggestive of a thermonuclear origin, although direct measurements of temperature and neutron spectrum are necessary to fully establish that the yield is predominantly thermonuclear in origin[46].  It is worth noting, that the neutron time-of-flight diagnostics showed a signal with the timing and magnitude that would be expected for D-T neutrons.  If this can be verified, it would indicate secondary fusion events from tritium produced in the D-D fuel, as had been shown on MagLIF experiments at Sandia.  In addition to the Zebra experiments, the team also completed a series of shots on the Cobra pulsed power machine at Cornell University for increased diagnostic access, especially for imagery to assess the stability of the inner surface of the liner during the implosion.   These experiments were not able to utilize D fuel, and thus did not offer insights on fusion performance.

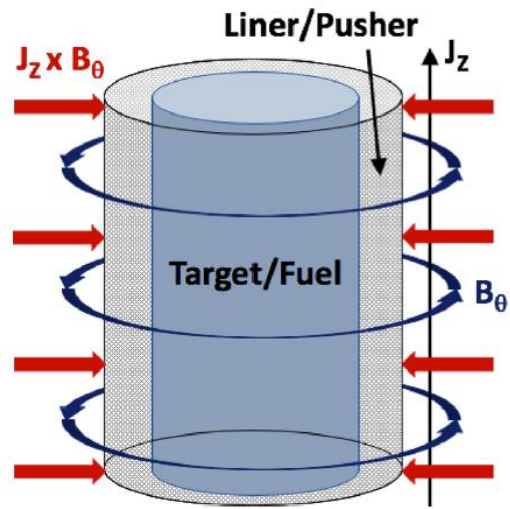

*Figure 4: Schematic of a liner-on-target Z-pinch, reproduced from Phys. Plasmas 26, 032708 (2019).*

Most of the simulations for the Staged Z Pinch were performed in MACH2, and as noted in the literature, there are disagreements over the extent of heating seen in MACH2 simulations of the Staged Z Pinch.  In particular, simulations in MHRDR show significantly lower levels of heating during the implosion, and suggest that the Staged Z Pinch will not extend to fusion breakeven[46].  The fusion community continues to investigate the origins of the major simulation discrepancies—to date no other teams have been able to reproduce the MIFTI team's simulation results.  Additional experiments with improved diagnostics for time-resolved data on temperature and density, and exploration of different implosion initial conditions, are needed to improve understanding of the Staged Z-Pinch[46] and its potential for scaling to net-gain fusion.

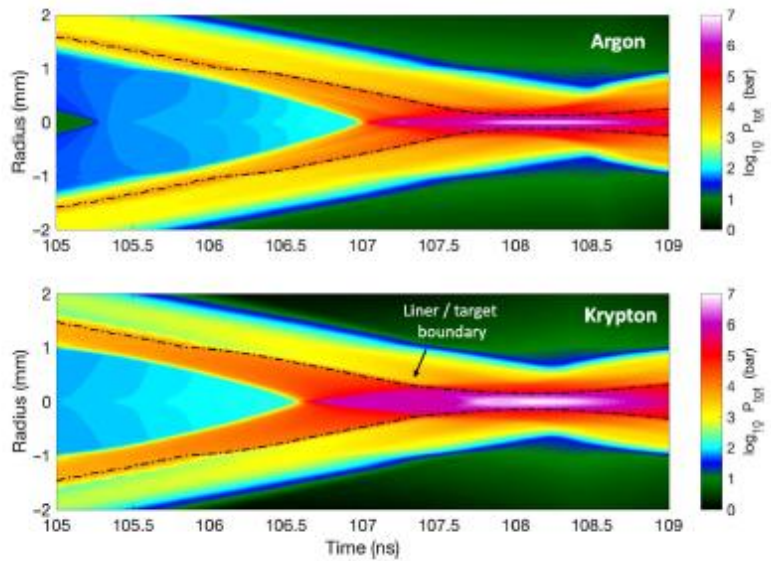

*Figure 5: Total plasma pressure in the staged Z-pinch as a function of time and radius near stagnation for Ar and Kr plasma shells compressing a deuterium target. Plots made with MACH2, a single-fluid radiation-MHD code.  Figure reproduced from Phys. Plasmas 26, 052706 (2019).*

*ii. Drivers*

*Los Alamos National Laboratory/HyperV Technologies: Plasma Liners For Fusion*

Los Alamos National Laboratory, along with HyperV Technologies and other partners, are developing a new MIF driver technology that is non-destructive, and should allow for more rapid experimentation and progress toward economical fusion power[47]. The team designed, built, tested, and deployed multiple plasma guns to produce hypersonic jets that merge to create a section of an imploding plasma liner, to support the development of the plasma-jet-driven magneto-inertial-fusion (PJMIF) concept [48][49]  Because the guns are located several meters away from the fusion burn region (i.e., they constitute a "standoff driver"), the plasma gun components should be protected from damage during repeated experiments. At the time of this writing, the team (which is continuing work through 2019 under a project extension) expects that by project completion they will better understand the behavior of plasma liners as they implode in order to demonstrate the validity of this driver design, optimize the precision and performance of the plasma guns, and obtain experimental data in a 36-gun experiment on ram-pressure scaling and liner uniformity critical to progress toward an economical fusion reactor.

 The project team designed, built, and tested seven state-of-the-art coaxial plasma guns, and used them to merge up to seven hypersonic plasma jets to form a section of a spherically imploding plasma liner [50] [48].  The team assessed two key scientific issues of plasma-liner formation via merging plasma jets: (i) shock heating leading to a degradation in the sonic Mach number of the merged jets, which would cause overly large spreading in the subsequently formed plasma liner leading to low calculated 1D energy gain [51] for the PJMIF concept, and (ii) degree of uniformity for the liner formed by discrete jet merging.  For (i), ion shock heating was measured in two- and three-jet merging experiments [52] , which benchmarked simulations showing that the liner-average Mach number remained above approximately 10.  For (ii), the first six-jet merging experiments were quite imbalanced due to large mass imbalance among the six jets[50].   An upgrade to the gas-valve design allowed for mass balance across seven jets of better than 2%.  The more balanced jets led to the formation of a section of the plasma liner in good agreement with simulations [48], [53]. Several upgrades to improve the gun precision, reproducibility, and maintainability were required and subsequently implemented, and testing is underway to qualify this second gun iteration to be the basis for a 36-gun, fully spherical plasma-liner-formation experiment. The additional gas-valve and gun-design iterations led to a substantial delay in the construction and fielding of the 36-gun experiment, which the team still hopes to execute.

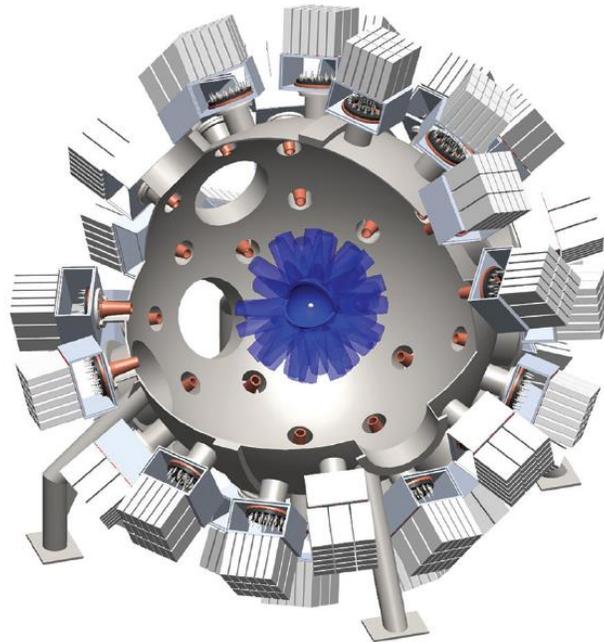

*Figure 6: Illustration of the planned Plasma Liner Experiment (PLX) set-up, which will ultimately have ≥36 coaxial plasma guns mounted around a 2.74-m vacuum chamber. Image from Scientia February 2017.*

If fully spherical plasma liners can be formed successfully, the key initial experiments would aim to characterize the peak ram-pressure scaling of the imploding plasma liner, as this is a key metric for the liner as a compression driver for MIF. Beyond that, the next priorities would be to (i) control and optimize the liner uniformity, (ii) initiate a program of PJMIF-compatible target formation, and (iii) compress the target using a plasma liner to show heating. An assessment of (ii) and (iii) are provided in [54]. In addition to the experimental work, the team has pursued modeling with team members University of Alabama in Huntsville and Brookhaven National Laboratory on the PLX experiments, and with Tech-X on plasma-liner compression of a magnetized target, with the objective on setting bounds on the minimum performance/uniformity of a liner and of target temperature and magnetization [55].

*NumerEx. Stabilized Liner Compressor For Low-Cost Fusion*

MIF covers a wide range of parameter space in ion density and magnetic fields in target plasmas, and the requirements in implosion velocity and pressure for a driver may vary greatly depending upon the the plasma parameters. The NumerEx team sought to develop a liquid metal liner implosion system for stabilized and repeatable compression at >1 km/s for magnetically confined plasmas at the lower end of ion density in the MIF regime. The driver concept is termed the "Stabilized Liner Compressor" (SLC), and it builds upon the "LINUS" approach developed at the Naval Research Laboratory in the 1970s[56][57]. The lower implosion velocity for the SLC as compared to MagLIF or PJMIF reflects the intended parameter space for liquid liner compression, which is designed for lower density magnetically confined plasmas, such as an FRC. (Where MagLIF or PJMIF cmight be considered as "ICF with greatly relaxed implosion velocities" due to reduced thermal losses with magnetic fields, the liquid liner approach might be considered as pulsed "MCF with a smaller reactor size" due to greatly increased ion density from pulsed compression). The SLC uses a rotating chamber in which liquid metal is formed into a hollow cylinder.[58] During compression the liquid will be pushed by pistons driven by high-pressure gas, collapsing the inner surface around a target on the axis. The rotation of the liquid liner maintains a smooth inner surface, and is intended to mitigate against Rayleigh-Taylor instabilities that would otherwise occur near peak compression of the plasma. During the ALPHA project, the NumerEx team made progress on the simulation and design of the SLC, with MACH2 results able to replicate experimental data from the prior LINUS system at NRL[9]. MACH2 also helped to establish a design basis for a 1 km/s implosion system with 5 cm bore diameter, NaK as the liquid metal, high pressure helium driving free annular pistons, and rotation about a central bearing[59]. The point design was for proof-of-concept on the SLC itself, compressing on an axial magnetic field in vacuum (and not pairing with a target plasma during the ALPHA project). In addition to the MACH2 simulations for implosion dynamics and magnetic compression, the design was qualified for safety margins and performance in ANSYS (a multi-physics simulation code) through the efforts of Applied Research Associates, a company that teamed with NumerEx in this project. The critical components in the design were a fast valve for <400-μs release of the >10-kpsi He plenum, and a triggering mechanism for reliable, synchronized firing of the pistons. The team was able to experimentally demonstrate a suitable fast valve design, but was not able to reach a qualified engineering design for triggering within the budget for the project, and it was determined to not yet proceed to build the proof of concept SLC demonstrator. With additional engineering and development for the triggering mechanism, a SLC system can be completed and demonstrated. In addition to the SLC design work, the NumerEx team also applied MACH2 for preliminary modeling of FRC injection and compression in the SLC system. Based upon FRC plasma parameters similar to those achieved in the Air Force Research Laboratory FRCHX Experiment[60], the NumerEx projections for the SLC show a maximum temperature of 2.4 keV <600-μs after compression begins, and a density of ~300x the initial density of a merged FRC target[61].

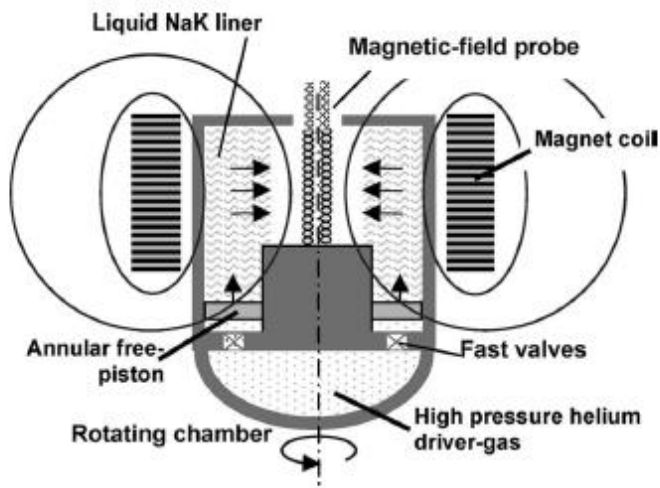

*Figure 7. Schematic of a stabilized liquid liner compressor. Image reproduced from IEEE Trans. Plasma Sci., vol. 36, no. 1, pp. 52–61, 2008.*

*iii. Applied Science of MIF*

*Sandia National Lab/University of Rochester Laboratory for Laser Energetics: Magnetization and Heating Tools for Low-Cost Fusion*

Sandia National Laboratories partnered with the Laboratory for Laser Energetics (LLE) at the University of Rochester to explore the behavior of magnetized plasmas under fusion conditions in the MIF regime, building upon the early successes of the MagLIF experiments at Sandia and seeking to collect more data, benchmark simulation codes, and establish a better understanding of MIF plasmas. A key challenge for MagLIF has been the low shot rate and limited diagnostic access on the Z Machine. To address this challenge, the Sandia/Rochester team collaborated in ALPHA to field the MagLIF concept on the University of Rochester LLE OMEGA facility to provide higher experimental throughput and diagnostic access to more rapidly explore the parameter space for MagLIF (and MIF more broadly). With some reconfiguration to deliver laser energy along an imploding cylindrical shell, the OMEGA facility is capable of conducting pre-heating, magnetization, and compression experiments that are similar to those possible on the Z-machine, though smaller in scale[62]. The Sandia/LLE team also collaborated to improve the performance of laser preheating in MagLIF, and thereby improve understanding of the initial conditions for the MagLIF experiments. The team was able to achieve both major goals for the project. In completing multiple rounds of integrated "mini-MagLIF" shots on OMEGA, the team was able to leverage an upgraded "MIFEDS" magnetization system (funded by DOE Office of Fusion Energy Sciences) to access magnetic fields of > 10 T (prior to implosion)[63]. While the smaller length scale of the mini-MagLIF experiment cannot match all of the conditions for the full scale MagLIF (for example, confinement of alphas and secondary tritium ions are significantly reduced in the smaller mini-MagLIF), the OMEGA experiments were able to confirm projections from LASNEX on fusion performance trends in the MagLIF configuration.[19] The team improved the uniformity and minimized the mix from the laser preheat in the ALPHA work, and fielded the improved preheat protocol in NNSA-funded MagLIF shots on the Z Machine. These new shots achieved a yield of 5e12, a 2x improvement in fusion yield over the prior record of 2e12 (DD neutrons) for MagLIF shots[11][19]. Future publications from the MagLIF team are anticipated to report another increase in DD yield, an improvement attributable, at least in part, to a laser pulse and experimental configuration (i.e. target density, preheat protocol) developed in ALPHA.

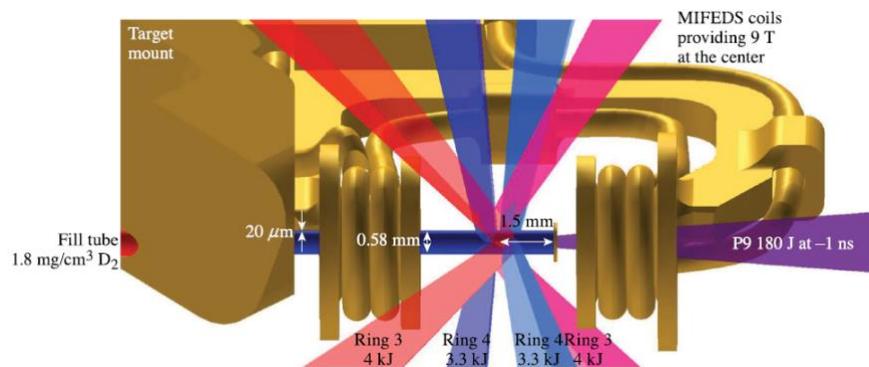

*Figure 8: Drawing of the OMEGA laser-driven MagLIF set-up, with only 8 out of 40 beams shown for clarity. Rings 3 and 4 are used for compression. Figure adapted from Phys. Plasmas 26, 022706 (2019)*

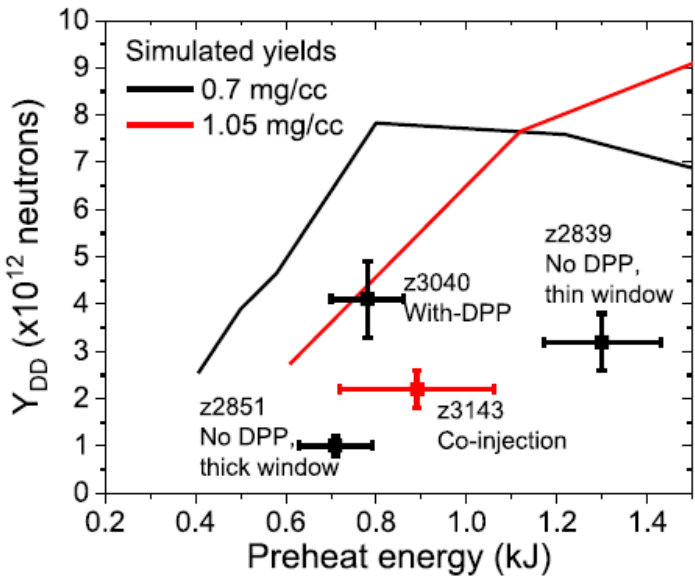

*Figure 9: Simulated curves of neutron yield vs. preheat energy, from 2D LASNEX simulations, compared to experimental result to determine the effect of a distributed phase plate (DPP) on yield. Plot reproduced from Phys. Plasmas 26, 032707 (2019).*

*California Institute of Technology/Los Alamos National Lab: Heating and Compression Mechanisms for Fusion*

As the ALPHA program sought to advance the science of magneto-inertial fusion with a limited budget, the use of low-cost experiments to develop an understanding of scaling in MIF plasmas was a critical focus. To this end, the California Institute of Technology (Caltech), in coordination with Los Alamos National Laboratory (LANL), investigated the scaling of adiabatic heating of plasma by propelling magnetized plasma jets into stationary heavy gas clouds to provide experimental data to investigate adiabatic compression. By moving to the liner's rest frame, the Caltech/LANL team was able to investigate the jet-target collision using many experiments with a wide range of parameters to determine the equation of state relating compression, change in magnetic field, and temperature increase. The experimental work was supplemented with advanced computer models[64].

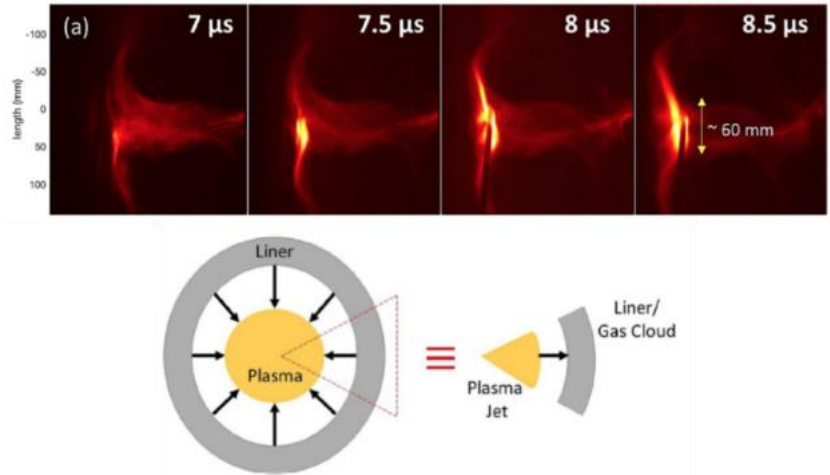

*Figure 10: Top: Image from a fast framing camera, showing time evolution of a collision between a plasma get and a gas target. Bottom Left: compression process for a generic MTF geometry. Bottom right: the same process in a reversed reference frame. Figure adapted from Phys. Plasmas 25, 112703 (2018)*

When a plasma is heated during compression the plasma density and temperature are expected to increase as $P \sim n^\gamma$ and $\sim n^{\gamma-1}$, where $P$ is plasma density and $T$ is plasma temperature. Gamma is the adiabatic constant, where $\gamma = (N+2)/N$ and $N$ is the number of degrees of freedom of an ion in the plasma, not the dimensionality of compression (which can take place in 1D, 2D, or 3D). If, for example, a plasma is sufficiently collisional when it is compressed in one or two dimensions $N$ will still be 3 because the collisional plasma leads to an equipartition of energy among each degree of freedom. For less collisional plasmas, even under a 3D compression $N$ may be less than 3. This project sought to determine $\gamma$ for the scaling of plasma heating during compression, via a shifted frame-of-reference experiment. The team found that density and magnetic field increased and jet velocity decreased during the compression. Interestingly, electron temperature featured a very complicated time dependence—temperature increased initially, and then dropped < 1 μs later (a time much shorter than the total compression time). The Caltech team attributed this to radiative loss due to three-body recombination in the plasma that results in hydrogen formation. The team found that a $\gamma$ of 5/3 (i.e. $N$=3, three degrees of freedom) was found to be a good fit for scaling this adiabatic compressionin 1D based on collisionality, with the necessity of correcting for electron radiative losses. Key results are summarized in [65]. Also notable, was that in order to make density measurements to verify the scaling relationship, the team had to develop and deploy a spatially translatable fiber-coupled interferometer, which is described in [66]. Finally, it is worth noting that the Caltech experiment is in a regime where three-body recombination is important, however magneto-inertial fusion experiments will not be, and thus more work needs to be done at higher densities to investigate adiabatic scaling at MIF-relevant conditions.

*iv. Exploratory Concepts*

*Lawrence Berkeley National Laboratory/Cornell University: MEMS Based Drivers For Fusion*

In this project the Lawrence Berkeley National Laboratory teamed with Cornell University to develop an ion beam technology that can be manufactured with low-cost, scalable methods. Ion beams have potential uses across the full range of fusion approaches, from ICF (with heavy or light ion fusion), MCF (with neutral beam heating), and MIF (with the "phi target" driven by ion beams) [67] [68][4]. However, currently available ion beam technology cannot practically or economically scale to the current densities and beam energies required in most fusion concepts (particularly those that demand the ion beam as the primary energy input). LBNL and Cornell developed a new ion beam architecture based on microelectromechanical systems (MEMS) technology that utilizes an array of "beamlets"—which will be scalable up to hundreds or thousands of beamlets per 4 to 12 inch wafer, and enabling very high system-level current density as the array of parallel beamlets deliver high ion flux even as no individual beamlet approaches space charge limits.

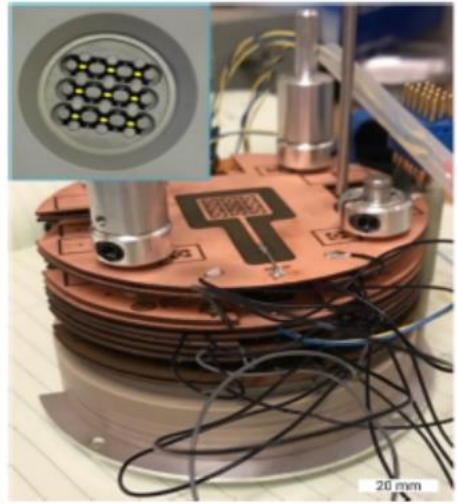
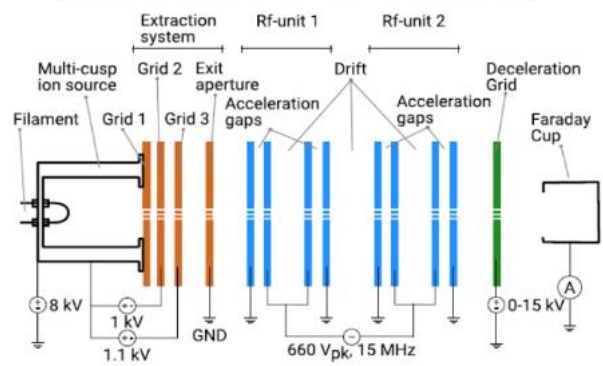

*Figure 11: Top: Photo of the final assembly, showing a downstream matching quadrupole (inset). Bottom: Experimental set-up for a filament driven ion souce to create a 3 x 3 array of beamlets. Figure adapted from Rev. Sci. Instrum. 89, 053302 (2018) and Rev. Sci. Instrum. 88, 063304 (2017).*

The design is based upon the MEQALAQ accelerator design developed at Brookhaven National Lab in the 1980's, but the addition of MEMS technology offers a path to mass manufacturability, and simplicity and scalability in the integration of RF with MEMS wafers[69]. The team was able to demonstrate proof of concept for the MEMS accelerator with a prototype compact accelerator fabricated from PCB board. The prototype demonstrated injection and transport of a 5-10 uA beam in a 3x3 beam array, and achieved ion acceleration of 0.5 kV/gap, for a gradient of about 0.3 MV/m[70], [71]. Building on these results, the team has used a compact, near-board RF driver to achieve up to 2.6 kV per gap[72]. While the performance of the prototype was relatively modest, the critical elements for the scalable MEMS design were established: a parallel array of beamlets, enabling high system-level current density, and the modular "stacking" of MEMS acceleration stages to enable high beam energy. The team has taken the first important steps to demonstrate that these elements can be "stacked" by demonstrating an acceleration of 125 V/gap at 22.7 MHz, and an increase of 250 V ion kinetic energy was observed in a two stage acceleration setup.[73] There is significant work remaining to mature and harden the design to develop practical accelerators, and the team is continuing to make progress on this front through continued ion beam development for the purposes of testing and characterization of materials for nuclear reactors and other potential applications in materials processing.[74]

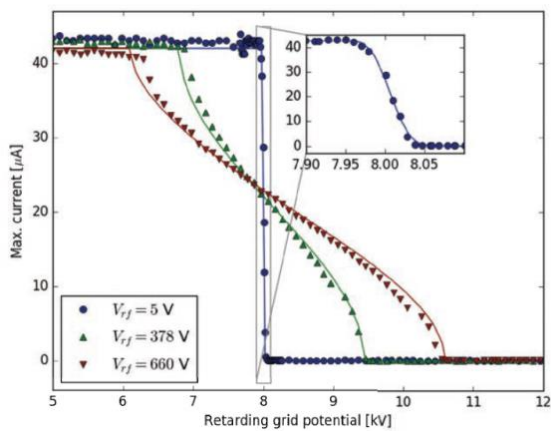

*Figure 12. Retarding-grid voltage scan with two RF units for low, medium and high RF voltages (markers) compared to simulations (solid lines). Inset is a scan with the RF signal generator at its lowest setting. Figure reproduced from Physics Procedia 90 (2017) 136 – 142.*

*Swarthmore College: Plasma Accelerator on the SSX*

This project sought to explore the properties of a fully relaxed parcel of magnetized plasma, known as a Taylor state. Taylor states are elongated structures with helical magnetic field lines resembling a rope. These Taylor states exhibit interesting and potentially very beneficial properties upon compression, and their long lifetimes may make them suitable as a fusion target if they are able to maintain their temperatures and stability long enough to be compressed to fusion conditions. The goal of Swarthmore was to explore whether the helical Taylor state could become a suitable target for compression. Under this project, the team sought to measure the equations of state for a Taylor state by carrying out compression and heating experiments by creating a parcel of magnetized plasma, compressing it against the end wall of a chamber, and measuring plasma density, temperature, and magnetic field during compression[75], [76]. The team conducted their experiments on the existing Swarthmore Spheromak Experiment device (SSX), which has an advanced diagnostic suite and the capability to perform 100 experiments per day, which enabled rapid progress in understanding the behavior of these plasma plumes and illuminating their potential for use as new targets in the pursuit of fusion reactors. By studying nearly 200 compression events the team was able to determine which equation of state best described the behavior of the Taylor states. At higher, MIF-relevant densities with higher collisionality, parallel and perpendicular equations of state may morph into the same. In this work the team found that the parallel component of the Chew, Goldberger, and Low (CGL) equation of state was the best fit for the behavior of the Taylor state.[75]  As a result, our understanding of the state of the art for Taylor state compression was advanced.  The Swarthmore team observed that Taylor states have very different flux loss behavior than other targets, like field-reversed configurations, which may lead to significant flux decay when translating a Taylor State through a series of coils[77]. This likely indicates that Taylor states may face significant challenges if considered as a "drop-in" replacement for FRCs in fusion systems. It remains inconclusive whther they can be a fusion target for other systems and more study is likely required. Beyond studying the behavior of a single Taylor state, work is ongoing to merge two helical Taylor-state plasmas to increase the density to values of interest (e.g., >$5e^{16}$/cc)  for subsequent liner compression.

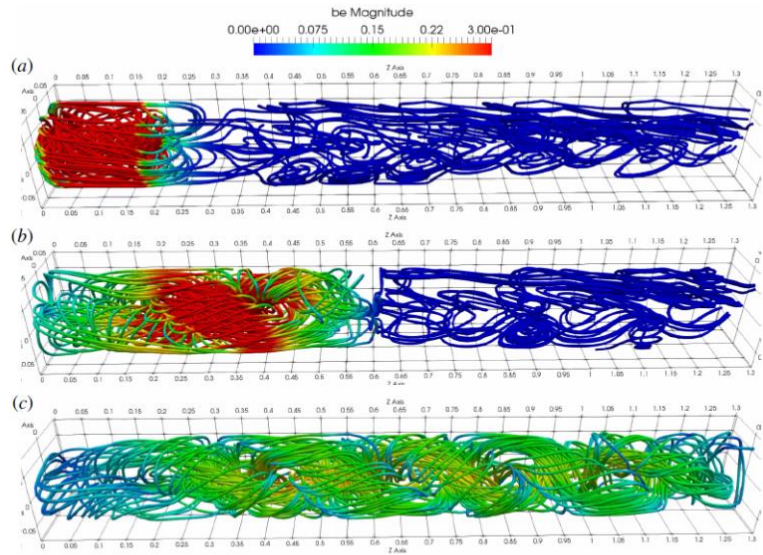

*Figure 13: Taylor State formation sequence. Top: spheromak formed at left. Middle: at 2 $\mu$s the spheromak tilts and begins to felax. Bottom: At 6.4 $\mu$s a relaxed, twisted Taylor state has formed. Reproduced from J. Plasma Phys. (2018), vol. 84, 905840614.*

### *Technology-2-Market (T2M) component of ALPHA*

ARPA-E programs are focused in scope and limited in time, and are by mandate not part of an existing roadmap within government or industry R&D.  As a result, navigating technology transition and securing follow-on funding is a key consideration for all ARPA-E programs from inception through project selection, program launch, and execution.  ARPA-E mandates that a modest fraction (5%) of awarded funding be applied toward technology transfer and outreach (TT&O) activities by each project team.  For the ALPHA program, the high technical risks and longer time horizons for fusion development raised unique questions and challenges for how to pursue the Technology-2-Market (T2M) component of a fusion program.  Beyond the TT&O activities pursued by each of the ALPHA teams for their projects, ARPA-E led coordinated, program-level T2M activities as part of ALPHA,  This section describes these efforts, which were intended to help build and sustain momentum for the lower-cost fusion energy concepts pursued in ALPHA.

A common theme in technology transition is to think from the end backwards.  At program launch, each ALPHA team was asked to prepare a T2M plan to describe how their concept could eventually help enable a more rapid, more affordable path to commercial fusion energy, and to identify the overall phases and estimated costs of development.  These T2M plans helped ARPA-E set end-of-project goals and expectations and understand the follow-on funding needs.

Early in the program, ARPA-E provided awardees with an overview of intellectual property (IP) considerations and commissioned a report on the fusion patent landscape [78].  From reviewing global patents in fusion energy, it was apparant that magnetic and inertial confinement approaches dominated the IP landscape; thus, perhaps magneto-inertial fusion is an area ripe for the creation of new IP.  The report also revealed that over half of the fusion energy IP assets had expired, primarily due to non-payment of fees.  As a result, ARPA-E suggested to ALPHA teams that they may wish to be judicious about the timing of filing given the 20 years of patent protection granted relative to the anticipated timeline to commercialization.

ARPA-E assisted the ALPHA teams in engagement with a broad group of stakeholders who will eventually all be needed to commercialize fusion energy, including (1) other government offices with interest in fusion, (2) several flavors of private investment (e.g. philanthropic, mission-focused, strategic, and traditional venture), as well as (3) representatives from the electric-power and power-plant industries. Representatives from each of these stakeholders were invited to speak and/or network at the ALPHA program annual review meetings, as well as at the annual ARPA-E Energy Innovation Summit.

To assist project teams in their engagements with potential investors in particular, ARPA-E commissioned a capital-cost study led by Bechtel National with support from Woodruff Scientific and Decysive Systems[79]. The Bechtel-led team assessed four conceptual 150-MWe fusion-power-plant designs based on concepts pursued by four of the ALPHA projects. The study arrived at overnight-capital-cost (OCC) estimates according to models developed for current nuclear power plant technologies (i.e., likely to be overly conservative), but, more importantly, performed a sensitivity analysis to look for key levers effecting the capital cost estimates. For example, the study indicated that the fusion core constituted less than half the OCC. Even using a conservative costing model, the study showed that ALPHA-supported concepts could potentially achieve OCC around $1B. An updated cost study will soon commence to arrive at more optimistic OCC estimates.

In 2018, ARPA-E commissioned an independent assessment of the prospects for low-cost fusion development by the JASON advisory group. The findings and recommendations can be found in the complete report[39]. In short, the report found that MIF is a plausible approach to controlled fusion and recommended further investment in MIF-relevant studies of plasma instabilities, transport, and plasma-liner interactions, as well as to focus on the near term goal of scientific breakeven in a system that plausibly scales to a commercial fusion power plant, although making fusion economically viable may be daunting. The study also recognized spin-off potential (e.g., fusion neutron source or fusion space propulsion) and that all promising approaches should be supported rather than focusing resources only on early front-runners.

Finally, as outreach to the scientific community, the ALPHA teams organized a miniconference in 2018 to report on their progress and results at the annual meeting of the American Physical Society's Division of Plasma Physics[80].

### *After ALPHA*

ALPHA was launched as a targeted exploration of a promising and underexplored space, with the particular goals of establishing new approaches that could be explored at lower costs (and correspondingly faster timelines) that could be compatible with private development. The results from the program show promise that intermediate-density approaches such as MIF and stabilized Z-pinches can produce thermonuclear plasmas of significant fusion yield in relatively small and inexpensive machines. However, the performance of these systems – in both Lawson triple product and fusion yield – remains several orders of magnitude below the best performance of high performance tokamaks or of NIF. The rapid progress for a modest investment in ALPHA is promising, but a great deal more development is required before any of these concepts can be established as viable candidates for fusion energy. In particular, the stabilized Z pinch is showing strong experimental evidence of achieving plasmas with both $T_e$ and $T_i$ exceeding 500 eV, which hopefully will be confirmed soon by direct diagnostic measurements[29]. In the sixty-plus years of controlled fusion research, only a very small handful of fusion configurations have exceeded this metric, typically at much greater investments.

These approaches do not represent the entirety of "non-traditional" fusion approaches that may offer attractive features for simpled reactor design and development. Many efforts outside of the ALPHA program, including TAE, General Fusion, and Tokamak Energy, referenced above, plus new entrants such as Commonwealth Fusion Systems all have significantly funded efforts towards compact or simplified reactor designs. ARPA-E supported work under the 2018 OPEN solicitation [81] to explore new concepts using novel RF heating for high-performance, compact FRCs by Princeton Fusion Systems [82]; and imposed dynamo current drive sustainment of a spheromak plasma by CTFusion[83]. University of Washington ALPHA spinout Zap Energy also received an OPEN 2018 award to continue increasing the electrical current and performance fo the stabilized Z pinch.

Significant progress in leveraging private industry for fusion development has also been made above and beyond the ALPHA program. Notably, between 2015 and 2018, the publicly disclosed private dollars invested in fusion doubled to over $1B. More than 10 private fusion development companies have garnered over $1M of private financing with more than half of these companies having raised over $10M. In addition, the Fusion Industry Association (FIA) was established in 2018 and today consists of nineteen member companies with additional associated affiliates[84]. The private companies recognize that fusion approaches with reduced cost, size, complexity, and eventual nameplate generation capacity (e.g. the maximum rated generation capacity) are needed to both accelerate development and, eventually, allow for market penetration[85]. This view was encapsulated in the recent U.S. National Academies report *A Strategic Plan for U.S. Burning Plasma Research* in the second of its two main recommendations: "the U.S. should start a national program of accompanying research and technology leading to the construction of a compact pilot plant that produces electricity from fusion at the lowest possible capital cost." [86]

Building on the ALPHA program and synergies with the FIA and the second recommendation of the National Academies report, ARPA-E is exploring opportunities for a potential new fusion program that is broader in scope than ALPHA while pursuing the same vision as ALPHA: catalyze R&D pathways to lower the cost and accelerate the development time scale for commercially viable fusion energy. In addition, the potential new program may place emphasis on achieving T2M outcomes that can help provide ARPA-E fusion awardees, private fusion ventures, and the larger fusion-energy R&D community a smoother and more sustainable development pathway toward commercially viable fusion energy that includes public, private, and philanthropic support and engagement.

The potential new program may seek to support efforts that:

- Advance the performance of innovative, lower-cost fusion concepts that have a plausible path toward timely, commercially viable fusion energy

- Catalyze development of enabling technologies (especially relating to the tritium fuel cycle and handling the extreme heat and particle flux from the fusion core) for commercially attractive fusion power plants with reduced size and nameplate generation capacity [87], leveraging the expertise and experience of R&D communities both within and beyond mainstream fusion

- Explore programmatic mechanisms to incentivize more cooperation between the public and private sectors, and maximize cost-effectiveness of fusion development, e.g., diagnostic resource teams[88]

- Pursue T2M and/or TEA (techno-economic analysis) activities that will help build the runway for fusion-energy development; examples could include (but are not limited to) conducting market analysis, supporting safety analysis to inform eventual regulatory decision-making, and educating/engaging the full energy ecosystem in appropriate ways (especially private investors, philanthropic foundations, and public-interest advocacy groups).

*Acknowledgements*

We acknowledge the ALPHA teams for their tremendous efforts in pursuit of lower-cost pathways to fusion energy, as well as the many visionary, brilliant, and difficult fusioneers who want to find a new way.